\documentclass[usegraphicx,usenatbib,useapjfonts,apj]{emulateapj}

\shorttitle{QSO candidates for SZ surveys}
\shortauthors{Jimenez et al.}

\begin{document}

\title{Southern Cosmology Survey III: QSO's from Combined GALEX and Optical Photometry}

\author{Raul Jimenez\altaffilmark{1,2}, David N. Spergel\altaffilmark{2}, Michael D. Niemack\altaffilmark{3}, Felipe Menanteau\altaffilmark{4}, John P. Hughes\altaffilmark{4}, Licia Verde\altaffilmark{1,2}, Arthur Kosowsky\altaffilmark{5}}
\altaffiltext{1}{ICREA \& Institute of Space Sciences (CSIC-IEEC), Campus UAB, Bellaterra 08193, Spain; \email{raulj@astro.princeton.edu}}
\altaffiltext{2}{Dept. of Astrophysical Sciences, Peyton Hall, Princeton University, Princeton, NJ-08544, USA}
\altaffiltext{3}{National Institute of Standards and Technology, Boulder, CO 80305, USA}
\altaffiltext{4}{Dept. of Physics \& Astronomy, Rutgers University, 136 Frelinghuysen Road, Piscataway, NJ-08854, USA}
\altaffiltext{5}{Dept. of Physics and Astronomy, University of Piitsburgh, Pittsburgh, PA 15208, USA}

\begin{abstract}
We present catalogs of QSO candidates selected using photometry from GALEX combined with SDSS in the Stripe 82 region and 
Blanco Cosmology Survey (BCS) near declination $-55$ degrees.  The SDSS region contains $\simeq 700$ objects with magnitude $i < 20$ and $\simeq 3600$ objects with $i < 21.5$ in a $\simeq 60$ square degree sky region, while the BCS region contains $\simeq 280$ objects with magnitude $i < 20$ and $\sim 2000$ objects with $i < 21.5$ for 
a $11$ square degree sky region that is being observed by three current microwave Sunyaev-Zeldovich surveys. Our QSO catalog is the first one in the BCS region. Deep GALEX exposures ( $ \gtrsim 2000$
seconds in $F_{UV}$ and $N_{UV}$, except in three fields) provide high signal-to-noise photometry in the GALEX bands ($F_{UV}, N_{UV} < 24.5$ mag). From this data, we
select QSO candidates using only GALEX and optical $r$-band photometry, using the method
given by Atlee and Gould (2008). In the Stripe 82
field,  60\% (30\%) of the GALEX selected QSO's with optical magnitude $i<20$ ($i<21.5$) also appear in the  Richards et al.\ (2008) QSO catalog constructed using 5-band optical SDSS photometry. 
Comparison with the same catalog by Richards et al. shows that the completeness of the sample is approximately 40\%(25\%). However, for regions of the sky with very low dust extinction, like the BCS 23hr field and the Stripe 82 between $0$ and $10$ degrees in RA, our completeness is close to 95\%, demonstrating that deep GALEX observations are almost as efficient as multi-wavelength observations at finding QSO's. GALEX observations thus provide a viable alternate route to QSO catalogs in sky regions where $u$-band optical photometry is not available. The full catalog is available at {\tt http://www.ice.csic.es/personal/jimenez/PHOTOZ}.
\end{abstract}


\section{Introduction}

With ground-based telescopes surveying the sky at millimeter wavelengths and angular resolutions of around 1 arcminute (APEX\footnote{ Atacama Pathfinder Experiment; www.apex-telescope.org}, SPT\footnote{South Pole Telescope; spt.uchicago.edu}, ACT\footnote{Atacama Cosmology Telescope; www.physics.princeton.edu/act}), and the the launch of the Planck satellite, we will be soon exploring how structures grew in the universe using the Sunyaev-Zeldovich effect \citep{SZ}. Indeed, the SPT collaboration has recently released its first results of their blind survey \citep{spt}. However, microwave fluctuation power spectrum at angular scales smaller than 4 arcminutes will be dominated by point source emission at all frequencies. It is therefore imperative to understand the point-source contamination of  the primordial and secondary power spectrum of the cosmic microwave background if we aim to extract cosmological information from small-scale microwave fluctuations. 

One of the major point-source contaminants is Quasars (QSOs). The Sloan Digital Sky Survey (SDSS) has revolutionized the QSO field by providing a photometric method to detect QSOs over large areas of the sky \citep{Richards,Richards2}. However, the SDSS only covers 25\% of sky, most of it in the northern hemisphere, and all of the ground-based high-resolution Sunyaev-Zeldovich experiments primarily observe
the southern hemisphere. Clearly a good photometric technique that relies on a small number of filters would be useful for identifying and removing QSO point sources. Because most future surveys will lack $u-$band photometry,\footnote{One notable exception is LSST (www.lsst.org), but science operations will not take place until 2016.} the SDSS technique cannot be used and one needs to rely on other photometric surveys.

Recently, \citet{Gould} proposed a method to discover QSOs using the Galaxy Evolution Explorer (GALEX) bands in combination with only one other photometric band (e.g., $r$). Because GALEX will eventually cover the whole sky, this provides the means to construct a catalog of QSOs in the regions covered by the future SZ experiments. This technique is similar to that proposed by \citet{Bianchi05,Bianchi07} but with the difference that Bianchi et al.\ exploit the superior photometric and morphological information of the SDSS sample, thus requiring costly multiwavelength photometry. Our GALEX Legacy program, part
of the Southern Cosmology Survey (SCS) program,  provides the largest continuous area and deepest coverage to date with overlapping optical data (the SDSS Stripe 82 field and Blanco Cosmology Survey 23-hour field). Exploiting this data set, we have used the \citet{Gould} technique to construct a QSO catalog in these two fields. We are the first ones to provide a QSO catalog for the BCS 23-hour field. All three ground-based microwave experiments are observing the 23-hour field, while ACT is observing Stripe 82. We have found in the SDSS Stripe 82 and BCS areas around 1000 (5500) QSO candidates for $i < 20$ ($i<21.5$). For the brighter sample, around $60\%$ of our QSO candidates are actual QSO's, as determined by comparison with SDSS-identified QSOs from the Richards et al. catalog, dropping to around $30$\% for the fainter sample. The sample is complete at the 40\% and 25\% level in the brighter and fainter samples, respectively, when compared to the QSOs in the Richards et al\ (2008) catalogue\footnote{{\tt www.physics.drexel.edu/$\sim$gtr/outgoing/nbckde/tab1.dat.bz2}}. Our completeness is significantly higher $\sim 95$\% for regions of the sky with very low dust extinction. We provide an electronic version of the QSO catalogs at {\tt http://www.ice.csic.es/personal/jimenez/PHOTOZ}.

\section{Input Catalogs}

Our GALEX observations comprise a Legacy program awarded in Cycle 3,
with the goal of mapping 100 deg$^2$ with 3 ks exposure time
per pointing in both the $F_{UV}$ and $N_{UV}$ filters. We chose to
map roughly 11 deg$^2$ covering the Blanco Cosmology
Survey\footnote{cosmology.uiuc.edu/BCS} (BCS) 23-hour field at
declination -55$^\circ$, and a larger area of the equatorial Stripe 82 
field covered by SDSS. Both areas have $griz$ optical observations,
and SDSS also has $u$ observations. 
We took advantage of the fact that the Stripe 82 survey area includes a
number of the GALEX Medium Imaging Survey (MIS) fields, which
already had many observations of 1.5 ks or longer, and therefore needed only
partial additional observations to reach our 3 ks target. When our program is finished, we will have collected around 210 ks of new GALEX observations.

Matches between the GALEX detections and SDSS or BCS data were done  by initially
assigning optical sources to a particular GALEX pointing if they fall within
$35.1'$ of the GALEX field center. This cuts the noisiest region of
the GALEX fields (near the edges), while maintaining complete sky
coverage between neighboring fields (i.e. leaving no gaps). Within
every GALEX field, each optical source is assigned a match to the nearest
GALEX object detected in the $N_{UV}$ band, if the GALEX object is 
within a $4''$ radius of the optical source; this
is a relatively conservative matching radius
\citep{Agueros_GALEX_SDSS_2005}. After all sources in the field are
assigned, the combined catalog is searched to test whether any two
optical sources are assigned to the same GALEX object. In the case of
overlapping assignments, the closest optical source to the GALEX
position is selected and the other is removed from the
catalog. Optical sources which do not have a GALEX detection are removed from the catalog.

\begin{figure}[ht!]
\includegraphics[width=\columnwidth,angle=0]{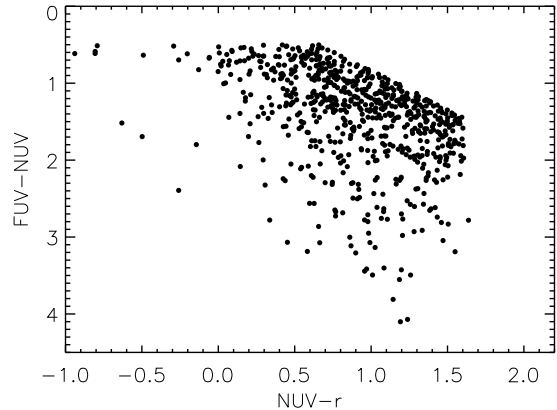}
\caption{GALEX-selected QSO candidates in the SDSS Stripe 82 sky region with magnitude $i \leq 20$, in a color-color plot  after \citet{Gould} cuts (see Table 1) have been applied. \label{f1}}
\end{figure}

\begin{figure}[ht!]
\includegraphics[width=\columnwidth,angle=0]{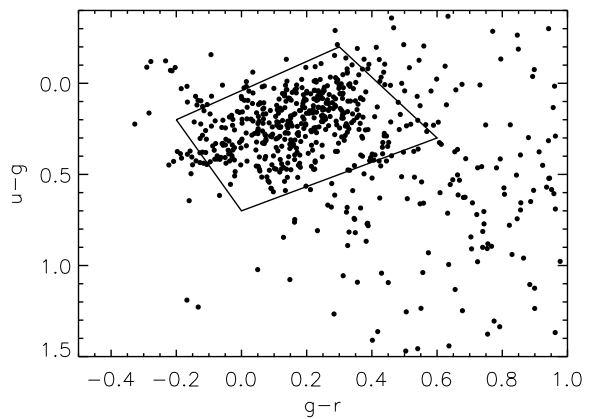}
\caption{QSO candidates from Fig.~\ref{f1} displayed in the  $u-g$ vs. $g-r$ plane. The area delineated by the solid lines is the \citet{Richards} most likely locus for QSO selection in the SDSS. Note that about 20\% of the GALEX-selected QSO candidates are outside the Richards et al.\ area. For the fainter sample, the plot is similar but with a larger number of outliers (around $40$\%). \label{f2}}
\end{figure}

\begin{figure}[ht!]
\includegraphics[width=\columnwidth,angle=0]{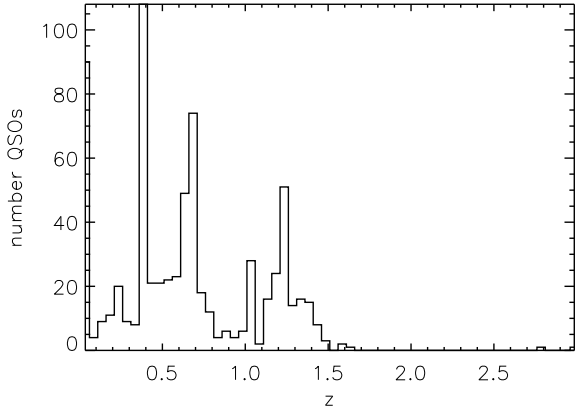}
\caption{Redshift distribution of QSO candidates in Fig.~\ref{f1}. Note that about 13\% of the candidates are flagged as stars, including most of the outliers in Fig.~\ref{f2}. \label{f3}}
\end{figure}

\begin{figure}[ht!]
\includegraphics[width=\columnwidth,angle=0]{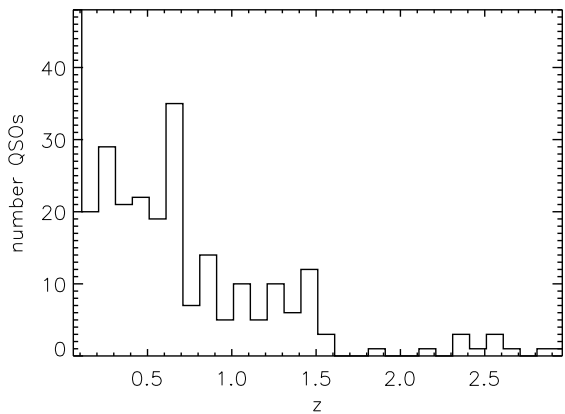}
\caption{Redshift distribution of GALEX-selected QSO candidates in the BCS 23-hour region with
magnitude $i < 20$. About 20\% of the candidates are flagged as stars. \label{f3b}}
\end{figure}

\begin{figure*}[ht!]
\includegraphics[width=2\columnwidth,angle=0]{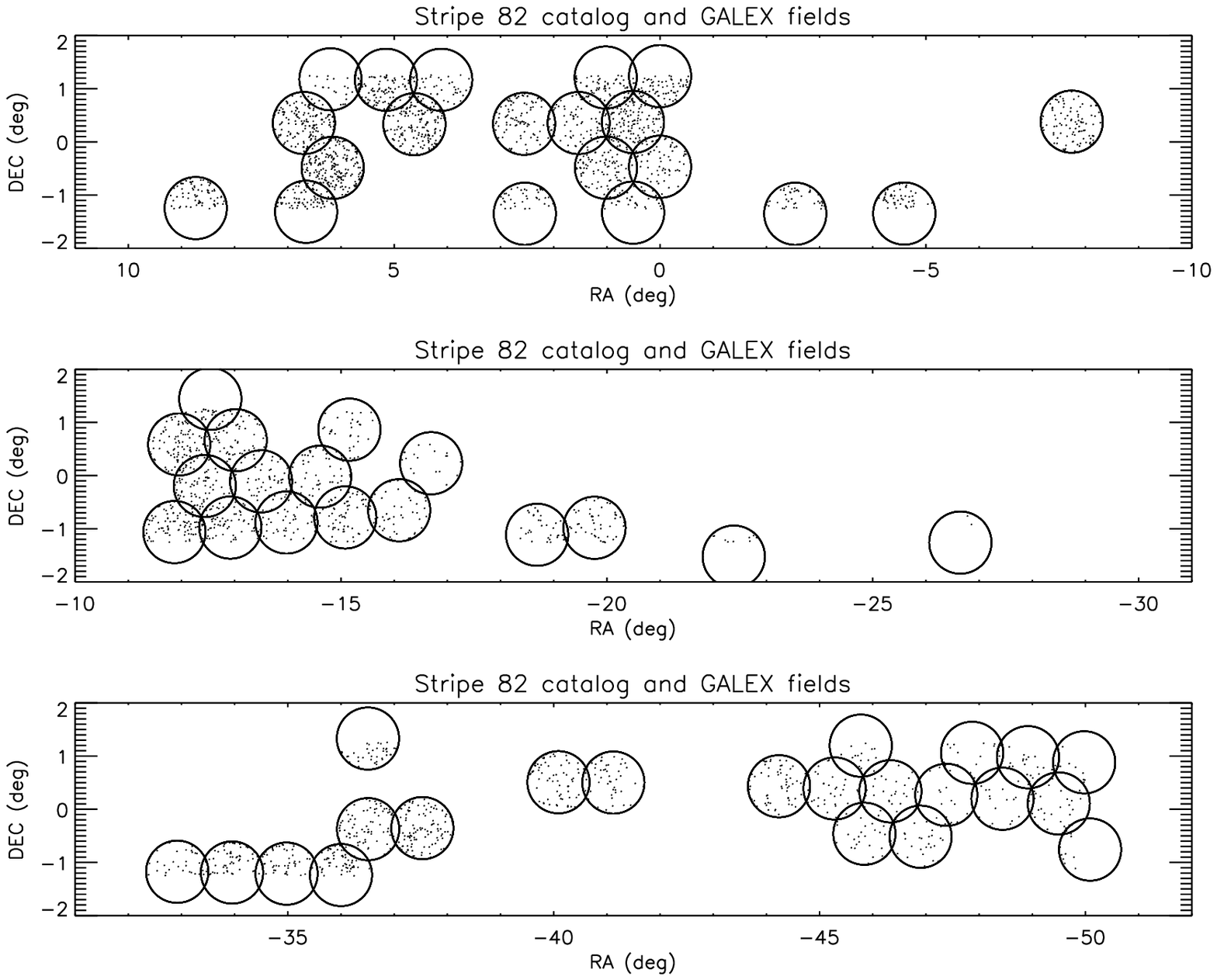}
\hspace*{4cm}
\includegraphics[width=\columnwidth,angle=0]{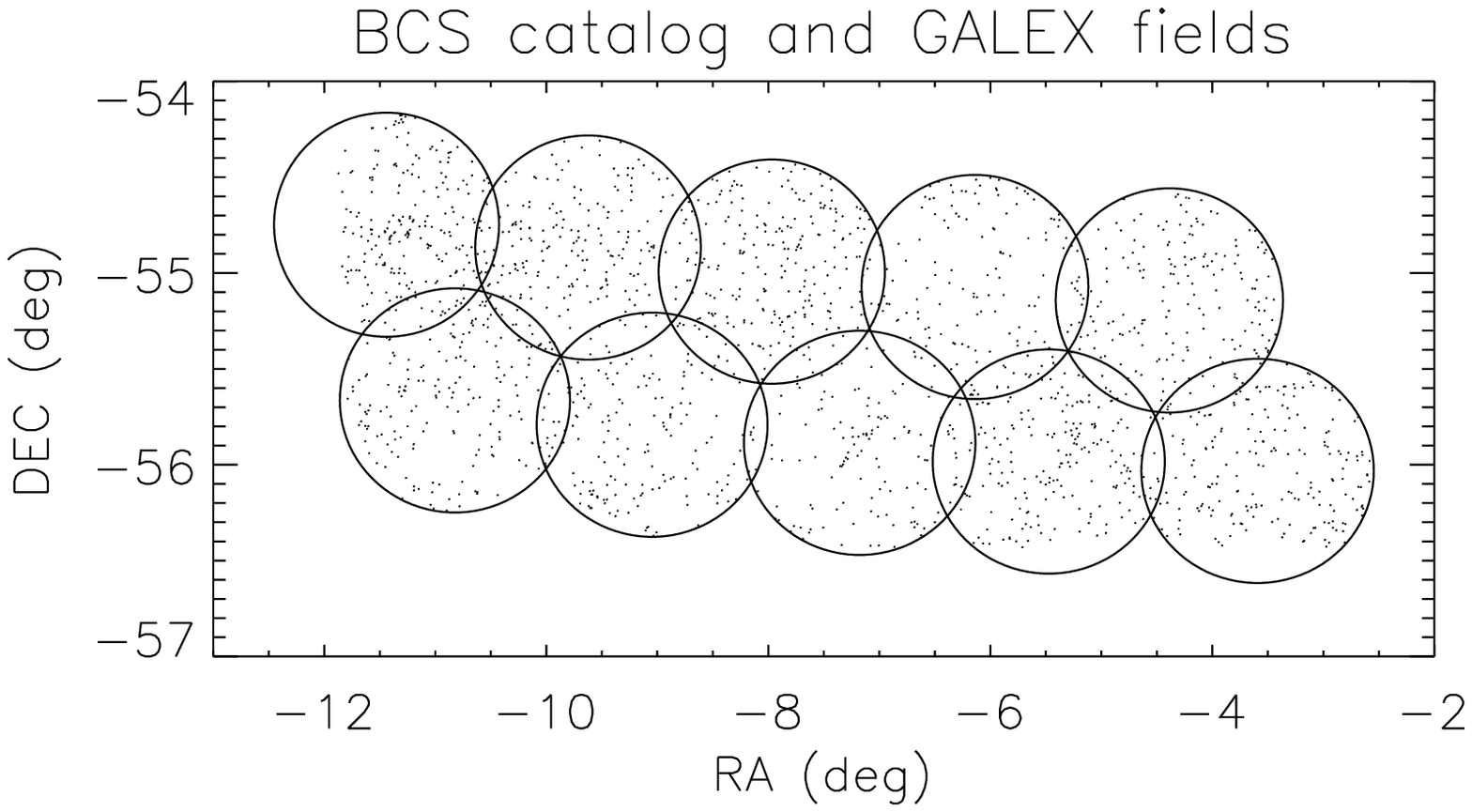}
\caption{QSO candidates location in the sky both in the Stripe-82 field (upper panel) and the BCS 23-hour field (bottom panel). Blank regions currently lack GALEX observations.
\label{f4}}
\end{figure*}

Because of the differences in the point spread functions  of
different instruments and between bands, simple aperture photometry
is not appropriate for this study. The SDSS point spread function widths are
approximately 1.5$''$ and vary with sky brightness
\citep{ABAZAJIAN_SDSS_2003}, while GALEX point spread function widths vary across the
field between roughly 4$''$ and 7$''$ \citep{Morrisey_GALEX_05}. Our
approach is to use AB magnitude measures that are as close as
possible to the total flux emitted by the galaxy in each band. For the SDSS data we use C-model magnitude measurements, which
consist of fitting models to a galaxy profile composed of
an exponential disc plus a deVaucouleurs profile; if an object is a point source, the exponential profile should dominate and give a good magnitude measurement. These fits are
integrated to three and seven times the characteristic radius, at
which point the function is truncated to smoothly go to zero. The
two fits are weighted based on the quality of the fit and combined
to obtain the best-fitting
profile.\footnote{www.sdss.org/dr5/algorithms/photometry.html} 
This measurement provides our estimate of the total photometry for
each SDSS band. 

For the BCS, we use the catalog developed by \citet{Menanteau08}. In this case the optical magnitudes 
are obtained using the {\tt{magauto}} feature in SExtractor \footnote{{\tt http://terapix.iap.fr}}, providing a good estimate of the total photometry for each BCS band. Note that  $r-$band magnitudes in BCS differ only by $\sim 0.03$ mag. from the SDSS $r-$band, which makes no difference in the QSO photometric selection. 

\section{Catalog construction and description}

\begin{deluxetable}{c}
\tablewidth{0pt}
\tablecaption{Color criteria from \citet{Gould} used in the present study}
\tablehead{\colhead{Boundary Criterion}} 
\startdata
FUV-NUV $\geq$ 37.314 (NUV-R) - 70.70372 \\
FUV-NUV $\geq$ (NUV-R) - 0.5 \\
NUV - R $\geq$ -0.895 \\
FUV-NUV $\geq$ 0.5 \\
FUV - NUV $\leq$ 4.343 \\
\enddata
\end{deluxetable}

We construct our catalog of QSO candidates by using the selection criteria of \citet{Gould} described in their Table~1 and Section~3, which we repeat here in our Table~1. Figure~\ref{f1} shows the QSO candidates in the $F_{UV}-N_{UV}$ vs. $N_{UV}-r$ color-color plot for the Stripe 82 region. After all selection cuts have been applied, we are left with a catalog of $\simeq 1000$ objects for a magnitude cut of $i < 20$. Because we have SDSS photometry for all candidates, we can compare with the catalogue from Richards et al.\ (2008). We match the two catalogues by finding matches within a 1.5" radius of each object in the Richards et al. catalog. We find that every object from our catalogue matches to only one Richards et al. object. As a function of magnitude we find that for $i < 19.1$ about 70\% of our QSO candidates are in the Richards et al. catalogue. This ratio decreases to $60$\% for $i < 20$ and $30$\% for $i < 21.5$. Using these numbers we conclude that the catalogs are complete with respect to the Richards et al. sample at the 40\% up to  $i < 20$ and  25\% for $i < 21.5$.  Because the GALEX observations are significantly influenced by diffuse dust extinction, we explore the completeness for regions in Stripe 82 with the lowest dust extinction. The lowest extinction region is in the range $0 <$ RA $ < 10$. In this region we find 95\% matches for $i < 21.5$ and a similar completeness level when  compared to the Richards et al. catalogue. This demonstrates that our technique is as successful as optical multi-wavelength searches in low extinction regions of the sky. 

To illustrate graphically the success of our scheme, we compare with the most likely color-color selection by \citet{Richards}, which has proven very successful at photometrically finding QSOs. This is shown in Fig.~\ref{f2}. The locus that \citet{Richards} use to select  QSO candidates is enclosed by the solid lines. Objects that fall within this region are more than 95\% likely to be true spectroscopically confirmed QSOs \citep{Richards}. For a magnitude cut of $i<  20$ (21.5) we find that only 20\% (40\%) of our objects lie outside the Richards-defined QSO region. Because we can compute accurate photometric redshifts using the GALEX and optical bands as done in \citet{Niemack}, we can test how many of our candidates are at $z=0$ and are therefore likely stars. We show the redshift distribution and spatial distribution of the QSO candidates in Fig.~\ref{f3} and \ref{f4}, for the Stripe 82 and BCS 23-hour regions respectively. Between 13\% and 20\% of the QSO candidates are likely stars; they all lie outside the Richards region. 

Our success rate at detecting QSOs is higher than \citet{Gould} because we have deeper GALEX exposures ($\gtrsim 2000$ seconds, except in three fields, which is $20$ times longer than the GALEX all-sky-survey they used), which allow for better sampling of the QSO emission features in the otherwise power-law-like UV spectrum of the QSO's.
In particular we find that at the bright end  ($i<19.1$) our technique finds 100\% of the QSOs of Richards  et al. catalog. For dimmer samples the success rate is lower, as more and more  QSO's are not detected in the GALEX bands, especially in the $F_{UV}$. We interpret this as an effect of extinction: the Richards et al.  QSO missing from the GALEX selection  are redder in the UV.

For the BCS sample we do not have $u$-band optical data, so we directly apply the cuts from Table~1 and present the redshift distribution and location of these host in Fig.~\ref{f3b} and \ref{f4}. As expected and discussed by \citet{Richards}, we are selecting QSOs at $z<2$. This can be seen in both Figs.~\ref{f3} and \ref{f3b}. The BCS catalogs contain $\sim 280$ objects with magnitude $i < 20$ and $\sim 2000$ objects with $i < 21.5$ for a $11$square degree region. The BCS 23-hour region is one of the lowest extinction areas in the sky, which allows for deeper GALEX observations than in the Stripe 82 and a higher rate (50-100\%) at finding QSO candidates. In fact, for the BCS 23-hour region we find $\sim 100$QSO's/sq. deg. which is similar to the QSO density found by Richards et al.

Table~2 shows the header of the catalog available online ({\tt www.ice.csic.es/personal/jimenez/PHOTOZ}) and describes the different entries for the catalog. We give the maximum likelihood redshift using the method described in Niemack et al. (2008) and the corresponding optical magnitudes in each survey. On the web page we provide catalogs for the two  different magnitude cuts ($i < 20$ and $i < 21.5$). These catalogs will be updated on-line as our GALEX observations are completed.

\begin{deluxetable}{cc}
\tablewidth{0pt}
\tablecaption{Description of columns in the SCS-QSO catalog. ML z refers to the best maximum lieklihood photo-z estimate as described in \citet{Niemack}. The catalog is available at {\tt http://www.ice.csic.es/personal/jimenez/PHOTOZ}.}
\tablehead{\colhead{Col. number}   &   {Description}}
\startdata
1  & SDSS-Cmodel u  \\
2  &  SDSS-Cmodel g | BCSmagauto g\\
3  &  SDSS-Cmodel r |  BCSmagauto r \\
4  & SDSS-Cmodel i  | BCSmagauto i\\
5  &  SDSS-Cmodel z | BCSmagauto z\\
6 & NUVmagauto  \\
7 &  FUVmagauto \\
8 &  ML z\\
9 &  GALEX   RA (deg)\\
10 &  GALEX   DEC (deg)\\
11 & GALEX objects within search radius\\
12 &  Optical galaxies matched to GALEX object\\
\enddata
\end{deluxetable}

\section{Conclusions}

We have constructed a new catalog of photometric QSO candidates from GALEX photometry and optical $r$-band data. In the SDSS Stripe 82
field, our selection criteria is successful in finding true QSO's at the 60\% (30\%) level with $i<20$ ($i<21.5$) when compared with the  Richards et al.\ (2008) QSO catalog constructed using 5-band optical SDSS photometry. Comparison with the same catalog by Richards et al. shows that the completeness of the sample is 40\%(25\%). For low extinction regions our completeness grows to 95\%.
This catalog covers some of the areas currently being scanned by microwave Sunyaev-Zeldovich experiments. It therefore provides a point source catalog to be masked out in these experiments. It can also be used to cross-correlate QSO positions with microwave temperature fluctuations to detect the Sunyaev-Zeldovich distortions due to the energy ejection from quasars
into the surrounding  intergalactic medium \citep{cha07,cha08,sca08}. Even at medium depth, GALEX coverage will have substantial utility in combination with optical imaging programs like Pan-STARRS (which does not have $u$ band coverage).
We anticipate that a GALEX observation program covering the SZ survey areas at medium depth in combination with single $r-$band photometry will be extremely useful to the SZ community. New QSO catalogs will also enable further studies including absolute astrometric reference frames. 

\acknowledgments
We thank Gordon Richards for useful comments. This work has been supported by GALEX grant GI3-095. The work of RJ is supported by grants from the Spanish Ministry for Science and Innovation and the European Union (FP7 program). RJ, DNS, JPH and FM are partially supported by NSF grant PIRE-0507768. LV acknowledges support by FP7-PEOPLE-2007-4-3-IRGn202182  and CSIC I3 grant 200750I034. AK was partly supported by
NSF grant AST-0546035.

The Galaxy Evolution Explorer (GALEX) is a NASA Small Explorer
(www.galex.caltech.edu). The mission was developed in cooperation
with the Centre National d'Etudes Spatiales of France and the Korean
Ministry of Science and Technology. We thank Joe Mohr and members of the Blanco Cosmology Survey team for obtaining optical imaging data used in this work.

\end{document}